\documentclass[letterpaper,prl,twocolumn,floatfix]{revtex4}  
\usepackage[normalem]{ulem}  
\usepackage{amsmath}
\usepackage[mathcal]{euscript}
\usepackage{mathrsfs}
\usepackage{float}
\usepackage[usenames,pdftex]{color} 
\usepackage{graphicx}     
\usepackage[letterpaper=true,bookmarks=false]{hyperref}           

\newcommand{\Tstr}{T_{\mathrm{string}}}
\newcommand{\scstr}{s_c^{\mathrm{string}}}
\newcommand{\Fdag}{F^{\ddagger}}
\newcommand{\Fin}{F_{\mathrm{in}}}
\newcommand{\deltaF}{\delta \! f}
\newcommand{\Nmax}{N_{\mathrm{max}}}
\newcommand{\partiald}[2]{\frac{\partial #1}{\partial #2}}
\newcommand{\Fbar}{\phi}

\begin{document}

\title{A universal origin for secondary relaxations in supercooled liquids and structural glasses}
\date{\today}
\author{Jacob D. Stevenson}
\affiliation{Department of Physics and Department of Chemistry and Biochemistry,
Center for Theoretical Biological Physics,
University of California, San Diego, La Jolla, CA 92093}
\author{Peter G. Wolynes}
\affiliation{Department of Physics and Department of Chemistry and Biochemistry,
Center for Theoretical Biological Physics,
University of California, San Diego, La Jolla, CA 92093}
\affiliation{e-mail: pwolynes@ucsd.edu}

\begin{abstract} 
  
  Nearly all glass forming liquids display secondary relaxations, dynamical
  modes seemingly distinct from the primary alpha relaxations.  We show that
  accounting for driving force fluctuations and the diversity of reconfiguring
  shapes in the random first order transition theory yields a low free energy
  tail on the activation barrier distribution which shares many of the features
  ascribed to secondary relaxations.  While primary relaxation takes place
  through activated events involving compact regions, secondary relaxation
  corresponding to the tail is governed by more ramified, string-like, or
  percolation-like clusters of particles.  These secondary relaxations merge
  with the primary relaxation peak becoming dominant near the dynamical
  crossover temperature $T_c$, where they smooth the transition between
  continuous dynamics described by mode-coupling theory and activated events.

\end{abstract}

\maketitle
\bibliographystyle{naturemagurl}

Diversity, a key feature of glassy systems, is most apparent in their
relaxation properties.  Dielectric, mechanical and calorimetric responses of
supercooled liquids are not single exponentials in time, but manifest a
distribution of relaxation times.  The typical relaxation time grows upon
cooling the liquid until it exceeds the preparation time, yielding a
non-equilibrium glass, which can still relax but in an age dependent fashion.
In addition to the main relaxations that are responsible for the glass
transition, supercooled liquids and structural glasses exhibit faster motions,
some distinct enough in time scale from the typical relaxation to be called
``secondary'' relaxation
processes\cite{adichtchev.2007,kudlik.1999,ngai.2004,wang.2007,lunkenheimer.2000}.
These faster motions account for only a fraction of the relaxation amplitude in
the liquid but become dominant features in the relaxation of otherwise frozen
glass, where they are important to the mechanical properties.  These secondary
relaxation processes in the solvation shell of proteins are also prominent in
protein dynamics\cite{frauenfelder.2009}.

The phenomenology of secondary relaxation has been much discussed but, owing
especially to the problem of how to subtract the main peak, the patterns
observed seem to be more complex and system specific than those for the main
glassy relaxation.  Some of the secondary relaxation motions are, doubtless,
chemically specific, occurring on the shortest length scales.  Nevertheless the
presence of secondary relaxation in glassy systems is nearly
universal\cite{thayyil.2008}.  In this paper we will show how secondary
relaxations naturally arise in the random first order transition (RFOT) theory
of glasses\cite{lubchenko.2007} and are predicted to scale in intensity and
frequency in a manner consistent with observation.

The RFOT theory is based on the notion that there is a diversity of locally
frozen free energy minima that can inter-convert via activated transitions.
The inter-conversions are driven by an extensive configurational entropy.  RFOT
theory accounts for the well known correlations between the primary relaxation
time scale in supercooled liquids and
thermodynamics\cite{xia.2000,stevenson.2005} as well as the aging behavior in
the glassy state\cite{lubchenko.2004}.  By taking account of local fluctuations
in the driving force, RFOT theory also gives a good account of the breadth of
the rate distribution of the main relaxation\cite{xia.2001,dzero.2008}.  Here
we will argue that RFOT theory suggests, universally, a secondary relaxation
also will appear and that its intensity and shape depends on the
configurational thermodynamics of the liquid.  This relaxation corresponds with
the low free energy tail of the activation barrier distribution.  The distinct
character of this tail comes about because the geometry of the reconfiguring
regions for low barrier transitions is different from that of those rearranging
regions responsible for the main relaxation.  Near to the laboratory $T_g$, the
primary relaxation process involves reconfiguring a rather compact cluster, but
the reconfiguring clusters become more ramified as the temperature is raised
and eventually resembling percolation clusters or strings near the dynamical
crossover to mode coupling behavior, identified with the onset of non-activated
motions\cite{stevenson.2006}.  Reconfiguration events of the more extended type
are more susceptible to fluctuations in the local driving force, even away from
the crossover.  These ramified or ``stringy'' reconfiguration events thus
dominate the low barrier tail of the activation energy distribution.  

When the shape distribution of reconfiguration processes is accounted for, a
simple statistical computation shows that a two peaked distribution of barriers
can arise.  This calculation motivates a more explicit but approximate theory
that gives analytical expressions for the distribution of relaxation times in
the tail.  In keeping with experiment, the theory predicts the secondary
relaxation motions are actually most numerous near the crossover, but of
course, merge in frequency with the main relaxation peak in time scale also at
that crossover.  Furthermore the relaxation time distribution for secondary
relaxations is predicted to be described by an asymptotic power law.  The
theory is easily extended to the aging regime where these secondary relaxations
can dominate the rearranging motions.

In RFOT theory, above the glass transition temperature,
the entropic advantage of exploring phase space, manifested as a driving force
for reconfiguration, is balanced by a mismatch energy at the interface between
adjacent metastable states.  For a flat interface in the deeply supercooled
regime the mismatch energy can be described as a surface tension that can be
estimated from the entropy cost of localizing a
bead\cite{kirkpatrick.1989,xia.2000}, giving a surface
tension $\sigma_0 = (3/4) k_B T r_0^{-2} \ln [1/(d_L^2 \pi e)]$ where $d_L$ is
the Lindemann length, the magnitude of particle fluctuations necessary to break
up a solid structure, and is nearly universally a tenth of the inter-particle
spacing, ($d_L = 0.1 r_0$).  The free energy profile for reconfiguration events
resembles nucleation theory at first order
transitions but is conceptually quite distinct.  Following
Stevenson-Schmalian-Wolynes (SSW)\cite{stevenson.2006} the free energy cost of an $N$
particle cluster with surface area $\Sigma$ making a structural transition to a
new metastable state may be written

\begin{equation}
  F(N, \Sigma ) = \Sigma \sigma_0 - N k_B T s_c  - k_B T 
  \ln \Omega(N, \Sigma) - \sum_{\textrm{particles}} \!\!\! \delta \! \tilde{f}  
  \label{eqn:full_profile}
\end{equation}

\noindent A key element of the free energy profile is the shape entropy $k_B
\log \Omega(N, \Sigma)$ which accounts for the number of distinct ways to
construct a cluster of $N$ particles having surface area $\Sigma$.  At one
extreme are compact, nearly spherical objects with shape entropy close to zero.
While objects such as percolation clusters or stringy chains have surface area
and shape entropy both of which grow linearly with $N$.  The last term of
equation \ref{eqn:full_profile} accounts for the inherent spatial fluctuations
in the disordered glassy system that give fluctuations in the driving force.
We presently ignore local fluctuations in the surface mismatch free energy, but
their inclusion would not qualitatively alter the
results\cite{biroli.2008,cammarota.2009,dzero.2008}.  We simplify by assuming
uncorrelated disorder, so each particle joining the reconfiguration event is
given a random energy, $\delta \tilde{f}$, drawn from a distribution of width
$\deltaF$.  The r.m.s.\ magnitude of the driving force fluctuations above $T_g$
follows from the configurational heat capacity through the relation $\deltaF
\approx T \sqrt{\Delta C_p k_B}$, a result expected for large enough regions.
We will assume no correlations for simplicity, but they can be included.

For nearly spherical reconfiguring regions forming compact clusters the shape
entropy is very small and the mismatch free energy is $\sigma_0 4 \pi (3 N
/(4\pi \rho_0))^{\theta/3}$ with $\theta = 2$ if fluctuations are small.  In
disordered systems the mismatch free energies grow with exponent $\theta$
generally less than 2 reflecting preferred growth in regions of favorable
energetics and the large number of metastable states which can wet the
interface and reduce the effective surface tension.  A renormalization group
treatment of the wetting effect\cite{kirkpatrick.1989} suggests that $\theta =
3/2$ in the vicinity of an ideal glass transition.  Incomplete wetting giving
strictly $\theta=2$ only asymptotically would not change the numerics of the
present theory much.  Whether complete wetting occurs for supercooled liquids
under laboratory conditions is still
debated\cite{capaccioli.2008,stevenson.2008a,cammarota.2009}.  The free energy
profile describing reconfiguration events restricted to compact clusters
becomes, then, $F_{\textrm{compact}}(N) = \sigma_0 4 \pi (3 N /(4\pi
\rho_0))^{\theta/3} - N T s_c$.  The minimum number of particles participating
in a reconfiguration event is determined by finding where the free energy
profile crosses zero.  For $\theta = 3/2$ the activation free energy barrier is
inversely proportional to the configurational entropy, leading to the
Adam-Gibbs\cite{adam.1965} relation for the most probable relaxation time
$\Fdag_{\alpha} / k_BT \sim \ln \tau_{\alpha} / \tau_0 \sim s_c^{-1}$.  Adding
fluctuations to the profile of compact reconfiguration events yields an
approximate Gaussian distribution of barriers with width scaling as
$\sqrt{N^{\ddagger}} \deltaF $.  Xia and Wolynes\cite{xia.2001}, and more
explicitly Bhattacharyya et al.\cite{bhattacharyya.2008}, have shown that, with
the inclusion of facilitation effects, the resulting barrier distribution
accounts for the stretching of the main relaxation process and yields good
estimates for how the stretching exponent varies with liquid
fragility.

\begin{figure*}[tp]
  \centering
  \includegraphics[width=0.8\textwidth]{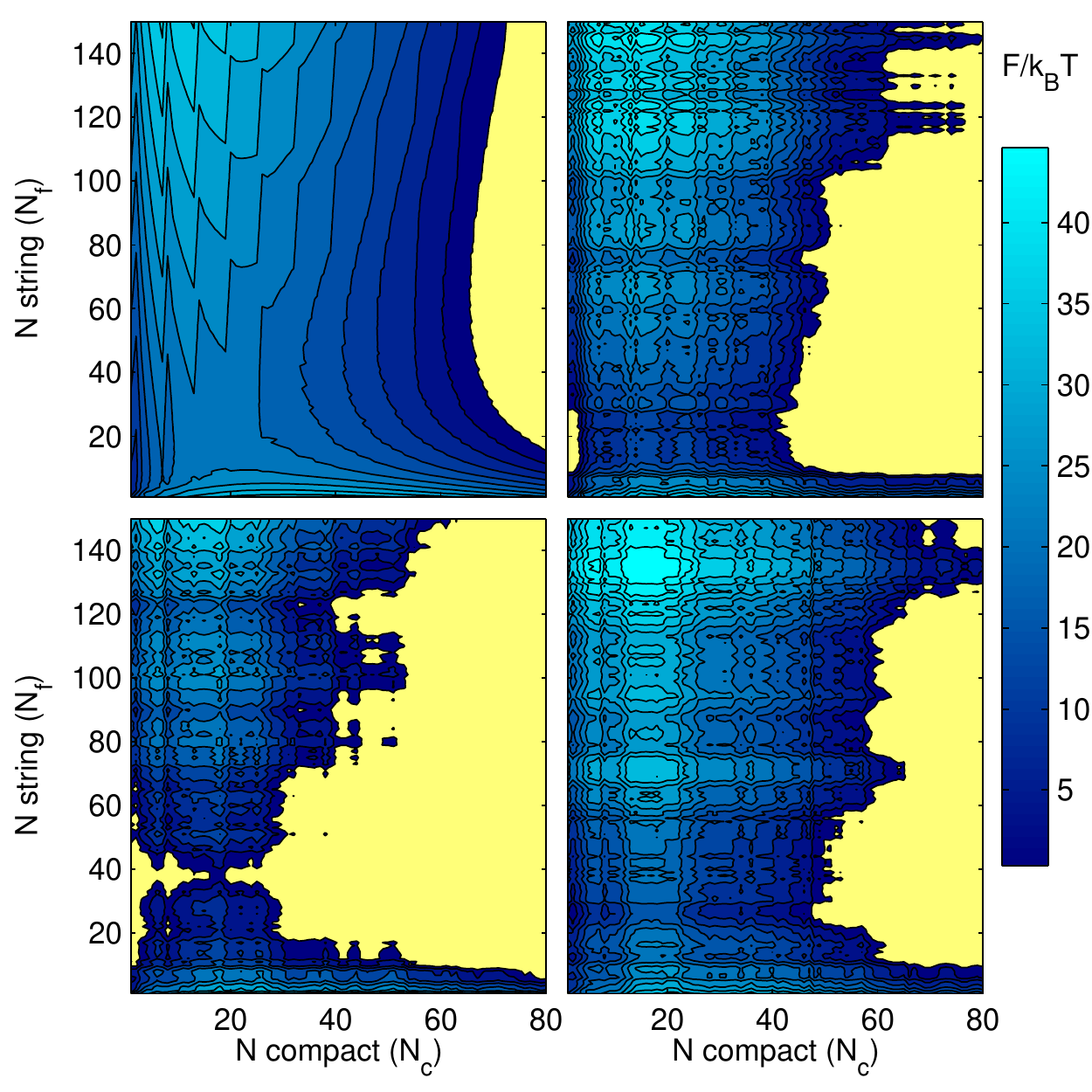} 
  \caption[The two dimensional free energy profile describing cooperative
  relaxation explicitly including fluctuations as a function of the number of
  compact particles $N_c$ and the number of stringy particles $N_f$ in the
  reconfiguring region.]{The two dimensional free energy profile describing
  cooperative relaxation as a function of the number of compact particles $N_c$
  and the number of stringy particles $N_f$ in the reconfiguring region.  The
  transition state separating the unreconfigured state ($N_c = N_f = 0$) with
  the final stable state (colored yellow in the figure) determines the free
  energy barrier to reconfiguration.  The upper left panel gives the profile
  ($s_c = 1.0k_B$) in the absence of fluctuations while the others demonstrate
  three possible realizations of fluctuations for a relatively strong liquid, a
  liquid having $\Delta C_P \approx 1 k_B$ per bead.  The fluctuations for the
  two dimensional profile are implemented as cumulative sums of local
  fluctuations in both $N_c$ and $N_f$.  For the situation described in the
  bottom right panel compact reconfiguration is required to overcome the free
  energy barrier. For the two other realizations the fluctuations are such that
  stable (yellow) regions exist along the vertical axis, meaning string-like
  reconfiguration is possible.  These stringy clusters, stabilized by
  fluctuations, account for the secondary relaxation.}
  \label{fig:2d_profile}
\end{figure*}

\begin{figure}[tp]
  \centering
  \includegraphics[width=0.48\textwidth]{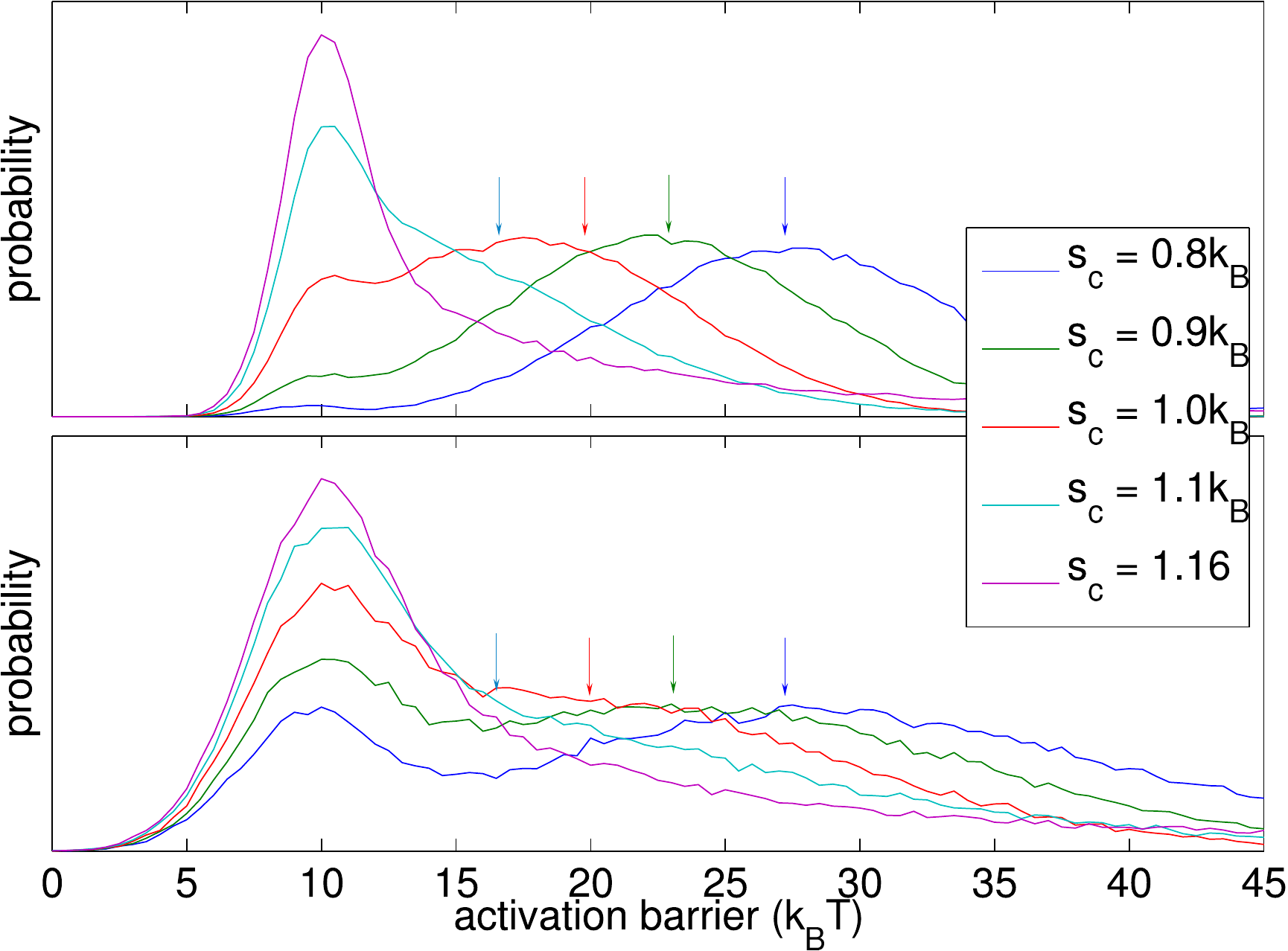}
  \caption[Probability distribution of free energy barriers governing
  relaxation events in supercooled liquids with fluctuations explicitly
  included.]{Probability distribution of free energy barriers governing
  relaxation events in supercooled liquids.  Different curves represent
  different temperatures, increasing from near the glass transition temperature
  to near the dynamical crossover temperature.  The arrows indicate the typical
  relaxation time predicted from the fuzzy sphere model without fluctuations.
  The dashed line gives the distribution of free energy barriers for a liquid
  just above the dynamical crossover temperature where primary relaxations
  disappear leaving only the secondary process. The top panel corresponds to a
  rather strong liquid with small fluctuations, $\Delta C_P \approx 1 k_B $ per
  bead --- a material similar to GeO$_2$. The bottom panel corresponds to a
  fragile liquid with larger fluctuations, $\Delta C_P = 3 k_B $ per bead --- a
  material similar to ortho-terphenyl. Secondary relaxations, i.e. relaxation
  events using string-like rearranging regions, increase in prominence as the
  temperature is increased, becoming the dominant process near the dynamical
  crossover temperature where the two peaks merge.  The units of the activation
  energy are given in $k_BT$, which assumes a mismatch penalty primarily
  entropic in nature, $\sigma_0 \sim k_BT$.  An energetic mismatch penalty,
  $\sigma_0 \sim k_BT_K$, would lead to Arrhenius behavior for the secondary
  relaxation process, as the distribution is peaked around the minimum free
  energy to initiate a stringy reconfiguration, $\Fin$.  In this calculation we
  have used the continuous approximation of $\Fin$ shown in the text.  The
  facilitation phenomenon, as described by Xia and
  Wolynes\protect{\cite{xia.2001}} and by Bhattacharya et
  al.\cite{bhattacharyya.2008} but not accounted for here, would shift weight
  from the largest free energies to the center of the primary peak, raising the
  overall height of the primary peak relative to the secondary peak.
  }
  \label{fig:fuzzy_prob}
\end{figure}

Restricting the reconfiguration events to
stringy clusters (using percolation clusters gives very similar results), gives a
free energy profile linear in the number of particles reconfigured, save for
the minimum cost $\Fin$ to begin to reconfigure a region: 

\begin{equation}
  F_{\textrm{string}}(N) = -N T (s_c - \scstr) + \Fin.  
  \label{eqn:string_profile}
\end{equation}

\noindent The critical ``entropy'' is given by $T \scstr = v_{int} (z-2) - k_B
T \ln (z-5) \approx 1.13 k_B T$.  This is the difference between the surface
energy written in terms of the coordination number of the random close packed
lattice $z \approx 12$ ($v_{int}$ is the surface tension per nearest neighbor)
and the shape entropy including excluded volume effects\cite{flory.1953}.  If a
bead can individually reconfigure then the cost to begin to reconfigure is
$\Fin = z v_{int} - T s_c \approx 2.5-2.9k_B T$.  If two must be moved then
$\Fin \approx 6.1k_B T$.  The continuous form of the surface mismatch energy
gives a somewhat higher value when applied to these small regions, giving
$\Fin^{\textrm{continuous}} = r_0^2 \sigma_04 \pi (3/(4 \pi))^{\theta/3} -Ts_c
\approx 10.5k_B T$ for a one particle reconfiguration.  The remarkably simple
free energy profile of equation \ref{eqn:string_profile} monotonically
increases, so that below $\Tstr$ (defined by $s_c ( \Tstr) = \scstr$)
reconfiguration via pure stringy objects is impossible. Above $\Tstr$ the same
process can occur with very small free energy barrier, having only to overcome
$\Fin$.  Thus $\Tstr$ signals the crossover from dynamics dominated by
activated events to dynamics dominated by non-activated processes.
Interestingly, this crossover is mathematically analogous to the Hagedorn
transition of particle theory\cite{hagedorn.1965}.  The predicted constant
value of $\scstr$ is confirmed experimentally\cite{stevenson.2006}.  In
contrast to the situation for compact reconfiguration, driving force
fluctuations dramatically alter the picture of stringy relaxation.  With
driving force fluctuations a lucky sequence of fluctuations can easily push the
nominally linearly increasing free energy profile to cross zero, so strings can
be active below $T_c$.

\begin{figure}[tp]
  \centering
  \includegraphics[width=0.48\textwidth]{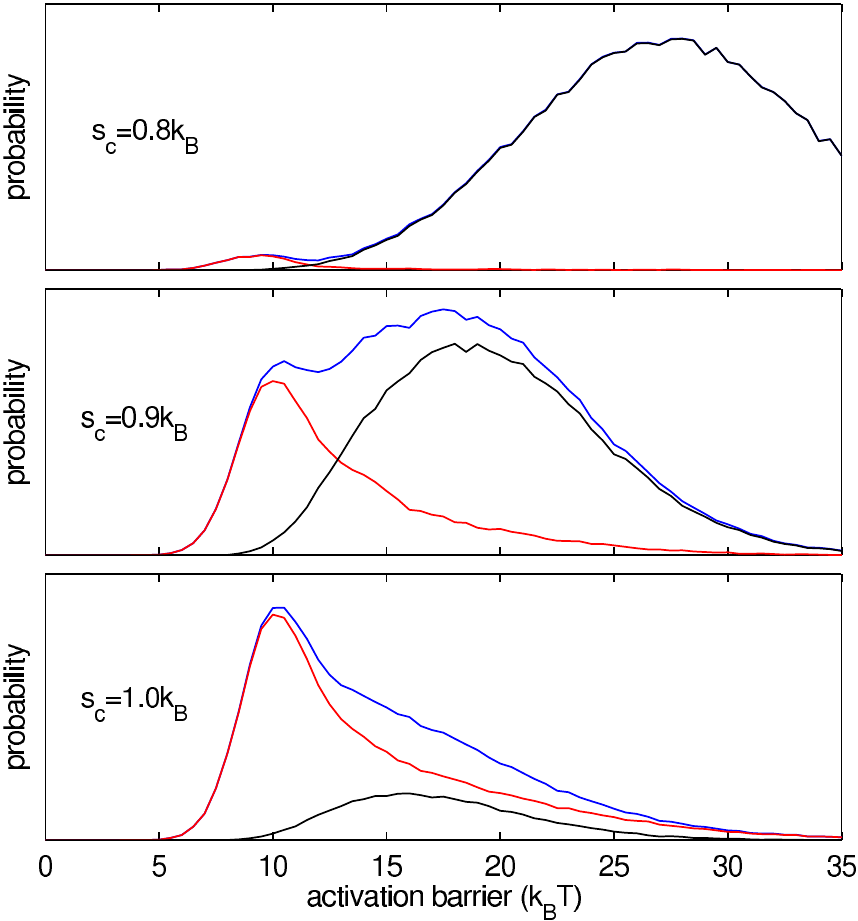}
  \caption{Distribution of free energy barriers for a strong liquid ($\Delta
  C_P \approx 1k_B$ per bead) separated into the contribution from secondary
  relaxations (red curves), corresponding to string-like reconfiguration as in
  panels b and c of figure \protect{\ref{fig:2d_profile}}, and primary relaxations (black
  curves), corresponding to compact reconfigurations.  The full distributions
  are given for comparison (blue curves).  The separation of the curves makes
  clear that as the dynamical crossover temperature is approached the primary
  relaxation becomes subordinate to the secondary relaxation.}
  \label{fig:separate_strong}
\end{figure}

\begin{figure}[tp]
  \centering
  \includegraphics[width=.48\textwidth]{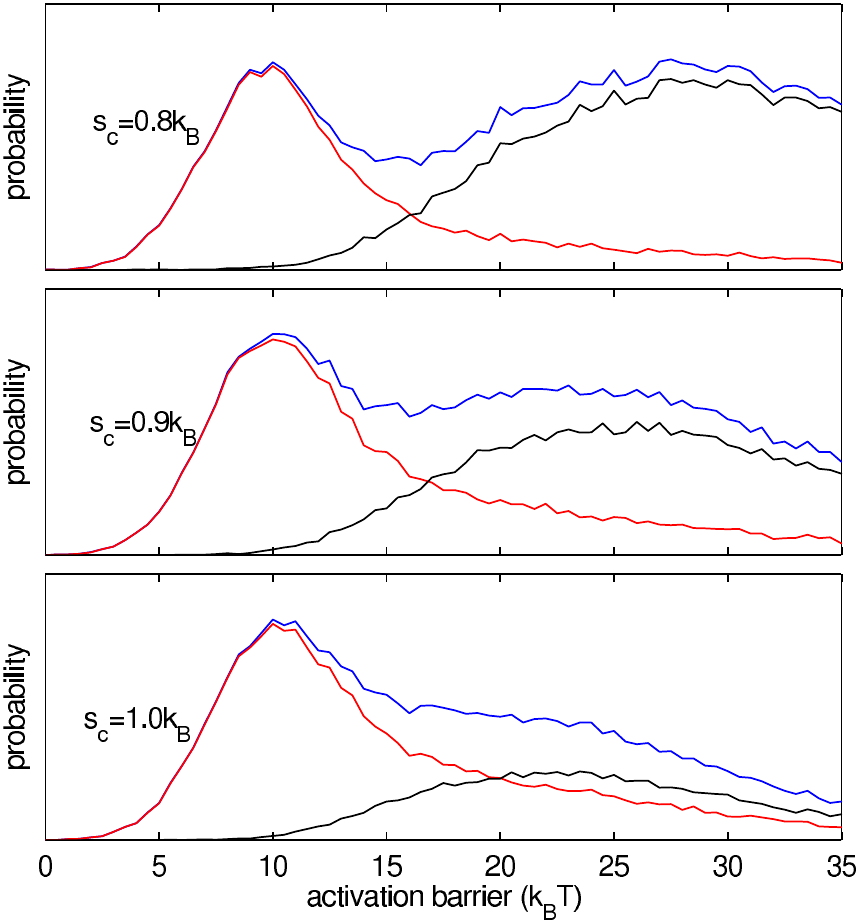}
  \caption{The corresponding results to those of figure
  \protect{\ref{fig:separate_strong}} but for a fragile liquid, one with
  $\Delta C_P \approx 3k_B$ per bead. }
  \label{fig:separate_fragile}
\end{figure}

\begin{figure}[t]
  \centering
  \includegraphics[width=0.48\textwidth]{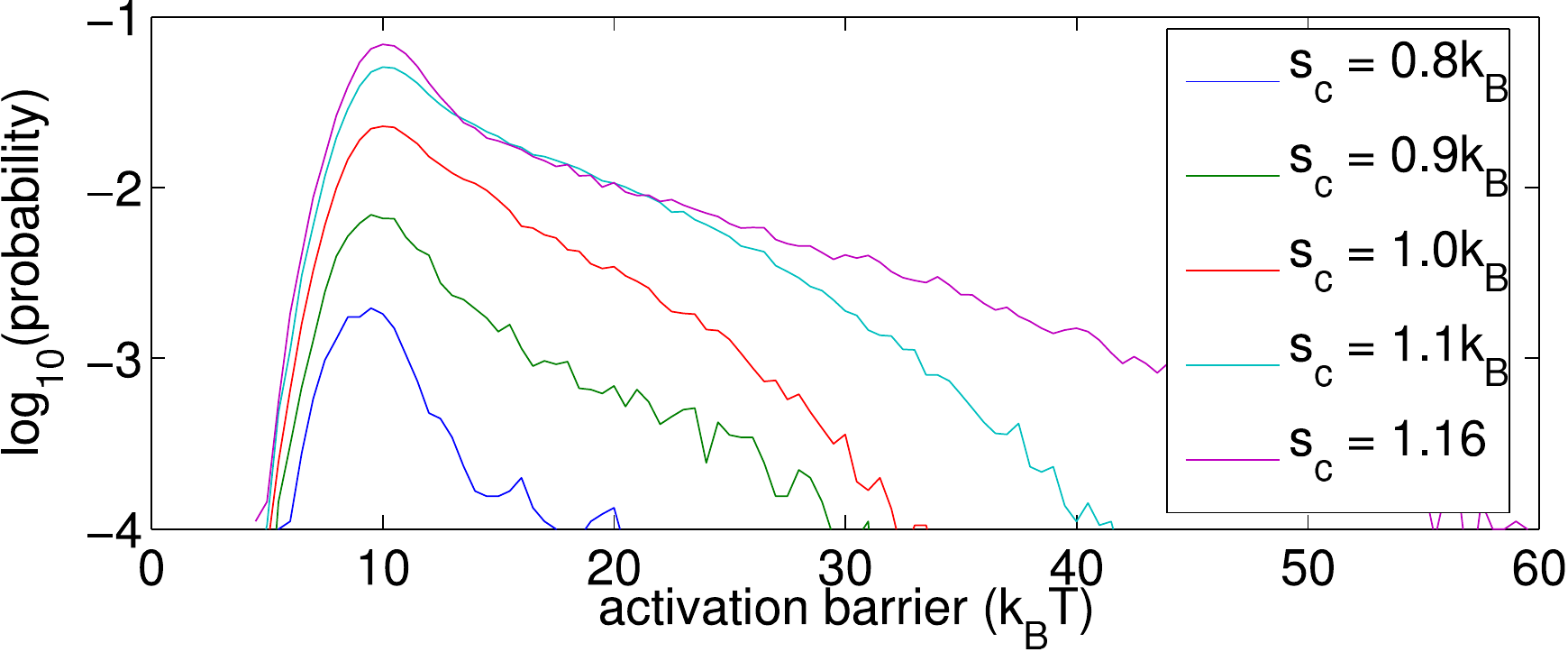}
  \caption{Distribution of free energy barriers for the secondary relaxation
  process from the statistical sampling of the fuzzy sphere model with
  fluctuations.  The data correspond to a strong liquid ($\Delta C_P \approx
  1$) and show that at higher temperatures (larger configurational entropies) the
  distribution decays more slowly.  This leads to wider activation energy
  distributions, matching the expectations of the analytical calculations.}
  \label{fig:decay}
\end{figure}

SSW\cite{stevenson.2006} introduced a crude model to estimate the shape entropy
and surface area of the range of shapes between the extremes.  This ``fuzzy
sphere'' model, consists of a compact center of $N_c$ particles with a stringy
halo of $N_f$ particles. With this interpolative model $T_g$ the preferred
shape of a reconfiguring region is largely compact, while the relevant regions
become more ramified close to $T_c$.  Fluctuations, aggregated cumulatively in
the compact core yielding a variance proportional to $N_c$, and cumulatively in
the stringy halo yielding a variance proportional to $N_f$, modify the SSW two
dimensional free energy landscape.  The local free energy plots for several
realizations of such accumulated fluctuations, assumed to have Gaussian
statistics, are given in figure \ref{fig:2d_profile}.  For some realizations
compact reconfiguration will still be required to overcome the free energy
barrier, but for other realizations the fluctuations are such that the free
energy cost for reconfiguration crosses zero along the $N_f$ axis (shown in the
figure in yellow) so the region is able to relax via a string-like
reconfiguration event.  These stringy rearranging clusters, stabilized by
disorder, we argue, are key contributors to the secondary relaxation process.
The statistics of the stable reconfiguration paths with lowest free energy
barrier are summarized in figure \ref{fig:fuzzy_prob} with barrier
distributions for two different values of $\deltaF$ corresponding to a strong
and a fragile liquid.  We can dissagregate these distributions into the parts
due, separately, to the compact events (``primary'') and to the string-like
fluctuation induced events (``secondary'').  These distributions are shown in
figures \ref{fig:separate_strong} and \ref{fig:separate_fragile}.  At
temperatures near $T_g$ compact primary relaxations dominate reconfiguration,
but as the temperature increases, fluctuations are able to stabilize
string-like reconfiguration more easily and the secondary relaxations increases
in prominence.  At the same time, with increasing temperature, the primary
relaxation peak shifts to lower free energy barriers, making it difficult to
distinguish between the two contributions.  As $T_c$ is approached and crossed,
the primary and secondary peaks merge with string-like reconfiguration clusters
becoming the dominant mode of relaxation.  For fragile liquids, i.e.  liquids
with larger configurational entropy fluctuations\cite{stevenson.2005}, the
secondary relaxation peak is generally more important than for strong liquids.
Both peaks are broader and begin to merge at lower temperatures for the more
fragile liquids.  Because facilitation effects are not explicitly accounted for
(these would mostly affect the higher barriers) it is not easy to directly
compare these predicted distributions with quantitative precision to
experiment.  In addition, the number of reconfiguring particles in the two
peaks is different, so their contributions to the measured amplitudes are
different as well.  Nevertheless, the predicted magnitude of the secondary
relaxation peak, as compared to the primary peak, seems to be somewhat larger
than experiments apparently show.  This disparity is more pronounced for
fragile materials.  The assumptions in the fuzzy sphere model, and especially
the assumption of uncorrelated disorder, apparently overestimate the influence
of the fluctuations, which are probably (anti-) correlated for the most fragile
systems.

The barrier distribution for reconfiguration events that take an ideal stringy
form can be explicitly calculated.  A similar analysis to the string case can
be applied to reconfiguration via percolation clusters but with somewhat
different numerical constants in the relation of the free energy profile to N.
This analytic calculation resembles that of Plotkin and Wolynes for the
``buffing'' of protein folding energy landscapes\cite{plotkin.2003}.  The key
to the calculation of $\Gamma(\Fdag)$, the distribution of free energy barriers
for events with less than $\Nmax$ displaced atoms, lies in mapping the problem
onto a random walk, or a diffusion process in free energy space.  Going to the
limit of continuous number of particles we may write a stochastic differential
equation for the free energy profile $d F / d N = d F_{\mathrm{string}} / d N +
\delta \tilde{f}$. The principal quantity to compute is $G( N, \Fdag; \Fin )$,
the probability that a reconfiguration path of $N$ particles has free energy
$F$ if the cost to initiate the reconfiguration event is $\Fin$.  The evolution
of $G$, that follows from the stochastic profile, is described by a diffusion
equation with drift subject to absorbing boundary conditions at both $F=0$ and
$F=\Fdag$.  These boundary conditions permit the calculation of the
distribution of free energy barriers by keeping track of the maximum excursion
of the random walk.

\begin{equation}
  \partiald{G}{N} + \Fbar \partiald{G}{F} = \frac{1}{2} \deltaF^2
  \partiald{^2 G}{F^2},
  \label{eqn:diffusion}
\end{equation}

\noindent The slope of the mean free energy profile $\Fbar = T(\scstr - s_c)$
depends simply on the proximity to the string transition.  $\Fbar$ is a string
tension reflecting the free energy cost of lengthening a string.  The
probability density for the maximum excursion of $F$ is then

\begin{equation}
  \Gamma ( \Fdag) = - \frac{\partial }{\partial \Fdag}
  \int_0^{N_{max}} \!\!\!\!\!\! d N \left< \frac{\deltaF^2}{2} \left.
  \frac{\partial G}{\partial F} \right|_{F=0} \right>_{0 < \Fin < \Fdag}.
  \label{eqn:gamma}
\end{equation}

\noindent The average $\langle \cdot \rangle _{0 < \Fin < \Fdag}$ is present to
integrate over the fluctuations in the free energy cost of initiating a string
--- capturing the statistics of the smallest possible activation barriers.  The
derivative with respect to $\Fdag$ converts from the cumulative probability
that the free energy barrier is below $\Fdag$ to the probability the free
energy barrier is between $\Fdag$ and $\Fdag + d \Fdag$.  $G$ can be calculated
explicitly by solving the diffusion equation using the method of images. The
result may be represented in closed form in terms of the Jacobi theta function,
however we leave the sum explicit to more easily examine the asymptotics

\begin{equation}
  \begin{split}
    G = & \frac{e^{\frac{\Fbar}{ \deltaF^2} (F-\Fin-\Fbar N/2)}}{\sqrt{2 \pi
    \deltaF^2 N}} \\ 
    \times &\sum_{n=-\infty}^{\infty} \bigg[  
    e^{-\frac{(2 n \Fdag + F - \Fin)^2}{2 \deltaF^2 N}} - e^{-\frac{(2 n \Fdag
    + F + \Fin)^2}{2 \deltaF^2 N}} \bigg]
  \end{split}
  \label{eqn:greens_final}
\end{equation}

In the integral of equation \ref{eqn:gamma} the cutoff $\Nmax$ reflects the
maximum size to which a stringy reconfiguration event would typically grow
before compact rearrangements dominate.  We estimate this maximum length as
$\Nmax \approx \Fdag_{\alpha} / \Fbar$, since certainly by that length the most
important reconfiguration events would be compact.  We can simplify by
smoothing the cutoff so that the finite integral $\int_0^{\Nmax} \!\! dN \cdot
$ is replaced by $\int_0^{\infty} \!\! dN \exp(-N/\Nmax) \cdot $.  The smoothed
distribution of free energy barriers follows directly from equations
\ref{eqn:gamma} and \ref{eqn:greens_final}

\begin{equation}
  \begin{split}
    \Gamma = &\partiald{ }{\Fdag} \Bigg< \exp\left( -\frac{\Fin
    \Fbar}{\deltaF^2} - \frac{1}{2} \Fin q \right) \\
    &\times
    \left(1- \frac{\exp(\Fin
    q)-1}{\exp(\Fdag q)-1 } \right) \Bigg>_{0<\Fin<\Fdag}\\ 
    &\textrm{where,  } q \equiv \frac{2}{\deltaF} \sqrt{ \frac{2 \Fbar
    }{F^{\ddagger}_{\alpha} } + \frac{\Fbar^2}{\deltaF^2}}
  \end{split}
  \label{eqn:gamma_solved}
\end{equation}

\noindent The result involves nothing more complicated then exponentials and
error functions.

The total magnitude of the secondary relaxation peak is estimated by
calculating the probability that fluctuations can stabilize a stringy
reconfiguration for any size barrier less than $\Fdag_{\alpha}$.  Integrating
$\Gamma$ over $\Fdag$ yields

\begin{equation}
    \Psi \approx \exp \left\{ -\frac{2\Fbar \left( \Fin - \Fbar \right)}{ \deltaF^2}
  \right\} 
  \label{}
\end{equation}

\noindent $\Psi$ increases with temperature as the dynamical crossover at
$\Tstr$ is approached, a trend that is validated
experimentally\cite{wiedersich.1999}.  At the crossover temperature and above,
this secondary relaxation becomes the only remaining mode of activated
relaxation.  The sharp transition from activated to non-activated motions at
$T_c$ that is predicted by the non-fluctuating RFOT theory, as well as by mode
coupling theory\cite{leutheusser.1984,gotze.1992} and the mean field theory of
supercooled liquids\cite{singh.1985,franz.2006,mezard.2000}, is therefore
smoothed out by the string-like activated events made possible by fluctuations
and exhibits no divergent critical behavior at $\scstr$\cite{bhattacharyya.2008}.  In this
temperature regime the secondary beta relaxations of mode coupling theory would
be present and overlap in frequency with the string-like activated secondary
relaxations, perhaps making the differentiation of the processes difficult.

$\Gamma$ can be approximated by the Gumbel extreme value distribution
function\cite{bertin.2005}.  For $\Fdag > \Fin$ the barrier distribution decays
exponentially as 

\begin{equation}
  \Gamma (\Fdag>\Fin) \sim \exp (- \Fdag q) 
  \approx \exp \left( -2
  \frac{\Fdag \Fbar}{\deltaF^2} \right)
  \label{eqn:decay}
\end{equation}

\noindent The results agree with the sampled distribution of barriers for the
string-like reconfiguration events alone (shown in figure \ref{fig:decay}).
Using the fact that $\tau = \tau_0 \exp(\Fdag/k_B T)$ equation \ref{eqn:decay}
gives a power law distribution of relaxation times $P(\tau) \sim
\tau^{-\gamma}$ where $\gamma \approx 2  (\scstr - s_c)/\Delta C_P + 1$. Well
above $T_g$ the high barrier side of the secondary relaxation blends in with
the primary relaxation peak.  Thus the secondary relaxation from ramified
reconfiguration events often appears as only a ``wing'' on the main
distribution\cite{blochowicz.2003,blochowicz.2006}.  

In the aging glass, the picture of secondary relaxation is slightly modified.
If the liquid fell out of equilibrium at $T_f$ then the frozen-in structure has
an average excess energy per particle $\epsilon (T_f) = \epsilon(T_K) +
\int_{T_K}^{T_f} dT \Delta C_P(T)$.  At temperatures $T<T_f$ a region of the
liquid undergoing reconfiguration would relax to a structure with average
energy $\epsilon (T) < \epsilon (T_f)$.  Thus the driving force for
reconfiguration gains an energetic contribution and the configuration entropy
in equation \ref{eqn:full_profile} is replaced by $Ts_c \to (Ts_c + \Delta
\epsilon)$ where $\Delta \epsilon  = \epsilon(T_f) - \epsilon(T) =
\int_{T}^{T_f} dT' \Delta C_P(T')$.  Lubchenko and Wolynes\cite{lubchenko.2004}
have shown that this additional driving force results in a change in slope of
the typical relaxation time as a function of temperature, and a transition to
nearly Arrhenius behavior as the system falls out of equilibrium.
Correspondingly, falling out of equilibrium causes the string tension to be
reduced by an amount $\Delta \epsilon$, giving $\Fbar = T\scstr - Ts_c - \Delta
\epsilon$ and making the system appear closer to the dynamical crossover than
an equilibrated system at the same temperature.  Furthermore, the driving force
fluctuations are frozen in as the aging glass falls out of equilibrium becoming
largely independent of temperature.   These changes broaden and flatten the
barrier distribution as the temperature is lowered, giving

\begin{equation}
  \Gamma (\Fdag>\Fin) \sim \exp \left( -2 \frac{\Fdag (T\scstr - Ts_c - \Delta
  \epsilon) }{T_f^2 k_B \Delta C_P(T_f) } \right)
\end{equation}

\noindent for large $\Fdag$.  The secondary relaxation strength in the aging
regime becomes

\begin{equation}
  \begin{split}
  \Psi \approx \exp \bigg\{ &-\frac{2(T\scstr - Ts_c - \Delta \epsilon) }{
  T_f^2 k_B \Delta C_P(T_f)} \\
  & \times \big( \Fin - (T\scstr - Ts_c - \Delta \epsilon) \big) \bigg\} .
  \end{split}
  \label{}
\end{equation}

\noindent In the limit $T \to 0$ the distribution of barriers becomes largely
independent of temperature with $\Gamma(\Fdag > \Fin) \sim exp(-\alpha \Fdag)$
and $\alpha \approx ((z-2) v_{int}(T_f) -\Delta \epsilon)/(T_f k_B \Delta
C_P(T_f))$.  For a broad enough distribution of barriers the dielectric
absorption spectrum is determined through the simple relation,
$\epsilon''(\omega) \sim P(\Fdag = -k_B T \ln \omega / \omega_0) \sim
\omega^{\alpha T}$, and becomes flat for low temperatures,
resembling the so called constant loss spectrum.  In a rejuvenating glass, an
aged system that is heated to a temperature above $T_f$, the energetic
contribution to the driving force is negative, $\Delta \epsilon < 0$.  In this
situation the system appears as if it is be further from the dynamical
crossover temperature than an equilibrated system at the same temperature and
secondary relaxations are relatively suppressed.

Nearly all glass forming liquids display, dynamical modes
seemingly distinct from the primary alpha relaxations.  We have shown that by
adding fluctuations to the existing structure of random first order transition
theory a tail develops on the low free energy side of the activation barrier
distribution which shares many of the observed features of the secondary relaxations.  The relaxation process responsible for the tail
differs from the primary relaxation mechanism in the geometry of the region
undergoing cooperative reconfiguration.  While primary relaxation takes place
through activated events involving compact regions, secondary
relaxation is governed by more ramified string-like, or percolation-like
clusters of particles.  While the existence of secondary relaxation is nearly
universal, the relevant motions are of shorter length scales than those for
primary relaxation, allowing additional material dependent effects and,
perhaps, less universal quantitative description than for the main relaxation.  The
present theory, however, suggests a universal mechanism for secondary
relaxation.  The theory points out some general trends about the way
these relaxations vary with temperature and substance which conform to
observation.

\bigskip 

\begin{acknowledgments} 
  Support from NSF grant CHE0317017 and NIH grant 5R01GM44557 is gratefully
  acknowledged.  Encouraging discussion on this topic with Vas Lubchenko, Hans
  Frauenfelder and J\"org Schmalian are gratefully acknowledged
\end{acknowledgments}


\end{document}